\documentclass[%
 reprint,
superscriptaddress,
 amsmath,amssymb,
 aps,
]{revtex4-2}

\usepackage{graphicx}% Include figure files
\usepackage{dcolumn}% Align table columns on decimal point
\usepackage{bm}% bold math
\usepackage{amsmath}% Better math
\usepackage{placeins}  % For FloatBarrier
\usepackage{mathtools} % For MoveEqLeft in split equation
\usepackage{xcolor}

\usepackage{float}

\begin{document}

\preprint{APS/123-QED}

\title{Deep Reinforcement Learning for Data-Driven Adaptive Scanning in Ptychography}

\author{M. Schloz }
\email[]{schlozma@hu-berlin.de}
\affiliation{%
Humboldt Universit{\"a}t zu Berlin, Institut f{\"u}r Physik \& IRIS Adlershof, Berlin, Germany
}%
\author{J. Müller}%
 \affiliation{%
Humboldt Universit{\"a}t zu Berlin, Institut f{\"u}r Physik \& IRIS Adlershof, Berlin, Germany
}%
\author{T. C. Pekin}%
 \affiliation{%
Humboldt Universit{\"a}t zu Berlin, Institut f{\"u}r Physik \& IRIS Adlershof, Berlin, Germany
}%
\author{W. Van den Broek}%
 \affiliation{%
Humboldt Universit{\"a}t zu Berlin, Institut f{\"u}r Physik \& IRIS Adlershof, Berlin, Germany
}%

\author{C. T. Koch}%
\affiliation{%
Humboldt Universit{\"a}t zu Berlin, Institut f{\"u}r Physik \& IRIS Adlershof, Berlin, Germany
}%

\date{\today}

\begin{abstract}
\noindent We present a method that lowers the dose required for a ptychographic reconstruction by adaptively scanning the specimen, thereby providing the required spatial information redundancy in the regions of highest importance. The proposed method is built upon a deep learning model that is trained by reinforcement learning (RL), using prior knowledge of the specimen structure from training data sets. We show that equivalent low-dose experiments using adaptive scanning outperform conventional ptychography experiments in terms of reconstruction resolution.

\end{abstract}

\maketitle

\section{Introduction}
\label{sec:level1}
Ptychography is a coherent diffractive imaging (CDI) method that has found use in light, x-ray and scanning transmission electron microscopies (STEM). The method combines whole diffraction patterns from spatially overlapping regions to reconstruct the structure of a specimen for arbitrarily large fields of view \cite{rodenburg2008ptychography}, with many advantages over other imaging methods \cite{humphry2012ptychographic,lozano2018low,jiang2018electron,zhou2020low}.  The development of new hardware \cite{mcmullan2016direct, tate2016high} and reconstruction algorithms \cite{thibault2008high, maiden2009improved} has led to ptychography becoming a mature electron microscopy technique \cite{jiang2018electron}. Current research to further improve this technique is driven by the desire to investigate thick samples \cite{van2012method,maiden2012ptychographic,van2013general,tsai2016x,jiang2018breaking} as well as to lower the required electron dose \cite{pelz2017bayesian, song2019atomic, schloz2020overcoming, chen2020mixed}. 

In order to lower the electron dose used, researchers have tried to vary various experimental parameters while preserving information redundancy through overlapping probes. One approach involves a defocused probe rastered across the specimen, with a less dense scan pattern. This uses therefore a lower dose than focused probe ptychography, but introduces additional complications for the reconstruction algorithm due to an increased need to account for partial spatial coherence in the illuminating probe \cite{chen2020mixed}. Another approach is simply to scan faster - by lowering the probe dwell time per probe position, an overall decrease in dose can be realized. However, this comes with its own limitations, as the physical limits of the electron source, microscope, and camera all must be considered. Finally, a third approach is the optimization of the scan pattern, deviating from a raster grid in favour of a generally more efficient pattern \cite{huang2014optimization}. This approach can, however, only yield a limited improvement in reconstruction quality as it is not capable of taking into account the structure of the specimen in the scan pattern.

In this paper we present an approach particularly tailored for electron ptychography that enables reduction of the electron dose through adaptive scanning. It is based upon the idea that, at atomic resolution, ptychography requires an increased information redundancy through overlapping illuminating beams only at regions that contain atomic structure of the scanned specimen. We present here an algorithm that scans only the regions with the highest information content in order to strongly improve the ptychographic reconstruction quality while keeping the total number of scan positions, and therefore the total dose, low. The scan positions are predicted sequentially during the experiment and the only information required for the prediction process is the diffraction data acquired at previous scan positions. 

The scan position prediction model of the algorithm is a mixture of deep learning models, and the model training is performed with both supervised and reinforcement learning. The synergy of deep learning and reinforcement learning has already shown strong performance in various dynamic decision making problems, such as playing Atari games \cite{mnih2013playing}, or Go \cite{silver2016mastering}, as well as tasks in robotics \cite{levine2016end, andrychowicz2020learning} and visual recognition \cite{sharma2015action, ba2014multiple, mnih2014recurrent}. The success of this approach, despite the complexity of the problems that they had to overcome, can be attributed to their algorithms' ability of learning independently from data. 

Similarly, the proposed algorithm here solves a sequential decision making problem by learning from a large amount of simulated or, if available, experimental ptychographic data consisting of hundreds to thousands of diffraction patterns. Here, the focus of the learning is specifically designed to maximize the dynamic range in the reconstruction for each individual scan position. The algorithm then transfers the learned behaviour it developed offline to a realistic experimental environment. 

Our approach is conceptually related to the subfield of computer vision that focuses on identifying relevant regions of images or video sequences for the purpose of classification or recognition. However, there are fundamental differences not only in the purpose, but also in the solution strategy for our application in contrast to computer vision tasks. Differences include a lack of direct access to images (updated real space information is only accessible through a highly optimized reconstruction algorithm), non-optimal parameter settings of the reconstruction algorithm and experimental uncertainties such as imprecise scan positioning of the microscope or contamination of the specimen requiring pre-processing of the reconstructed image, and the necessity of a much larger number of measurements requiring methods that improve the performance of the sequential decision making process. 

Work in adaptive scanning for x-ray fluorescence imaging \cite{betterton2020reinforcement} and for scanning probe microscopy \cite{vasudevan2020autonomous} has recently been reported. The work in \cite{betterton2020reinforcement} uses RL to determine the exposure time on a per pixel basis sequentially for multiple apertures that vary in their respective resolution. It is therefore more closely related to previous work in scanning electron microscopy that divides the measurement into a low-dose raster scan and a subsequent high-dose adaptive scan \cite{dahmen2016feature}. The latter work in \cite{vasudevan2020autonomous} uses Gaussian processes based Bayesian optimization to sequentially explore the image space with the scanning probe. However, it has been reported that this model suffers in performance as it lacks prior knowledge of the domain structure, which can be compensated by including a deep learning model with domain specific knowledge. Our proposed algorithm is the first application of adaptive scanning to ptychography, and is further unique in that the scan pattern is predicted using prior knowledge about the sample in the form of a pre-trained deep RL network, thereby improving performance. Our research forms a basis for a new avenue of automated and autonomous microscopy \cite{kalinin2021automated}.

With the ever-increasing data storage capacities, implementations of data infrastructures and data sharing platforms \cite{o2016materials, draxl2018nomad, draxl2019nomad,dillen2019zenodo}, access to ptychographic data will be further facilitated and data-driven adaptive scanning schemes can be applied to a vast number of ptychographic experiments.

We demonstrate the performance of our algorithm using experimentally acquired data. Our analysis shows that the algorithm can sufficiently learn information about the structure of a material from data in order to optimize the scan behaviour of the microscope in a real experiment. For low dose experiments, we show that adaptive scanning can improve the ptychographic reconstruction quality by up to $25.75\%$ and the resolution by up to $31.59\%$ compared to a non-adaptive (random) scan method. Adaptive scanning allows for the retrieval of the material's structure in this low dose regime and even improves the resolution of the reconstruction when compared to the reconstruction obtained using the conventional high dose raster grid scan approach. 

\section{Methods}
\subsection{Image formation in ptychography}
\label{sec:ptycho}
\begin{figure}%[H]
\includegraphics[width=0.9\linewidth]{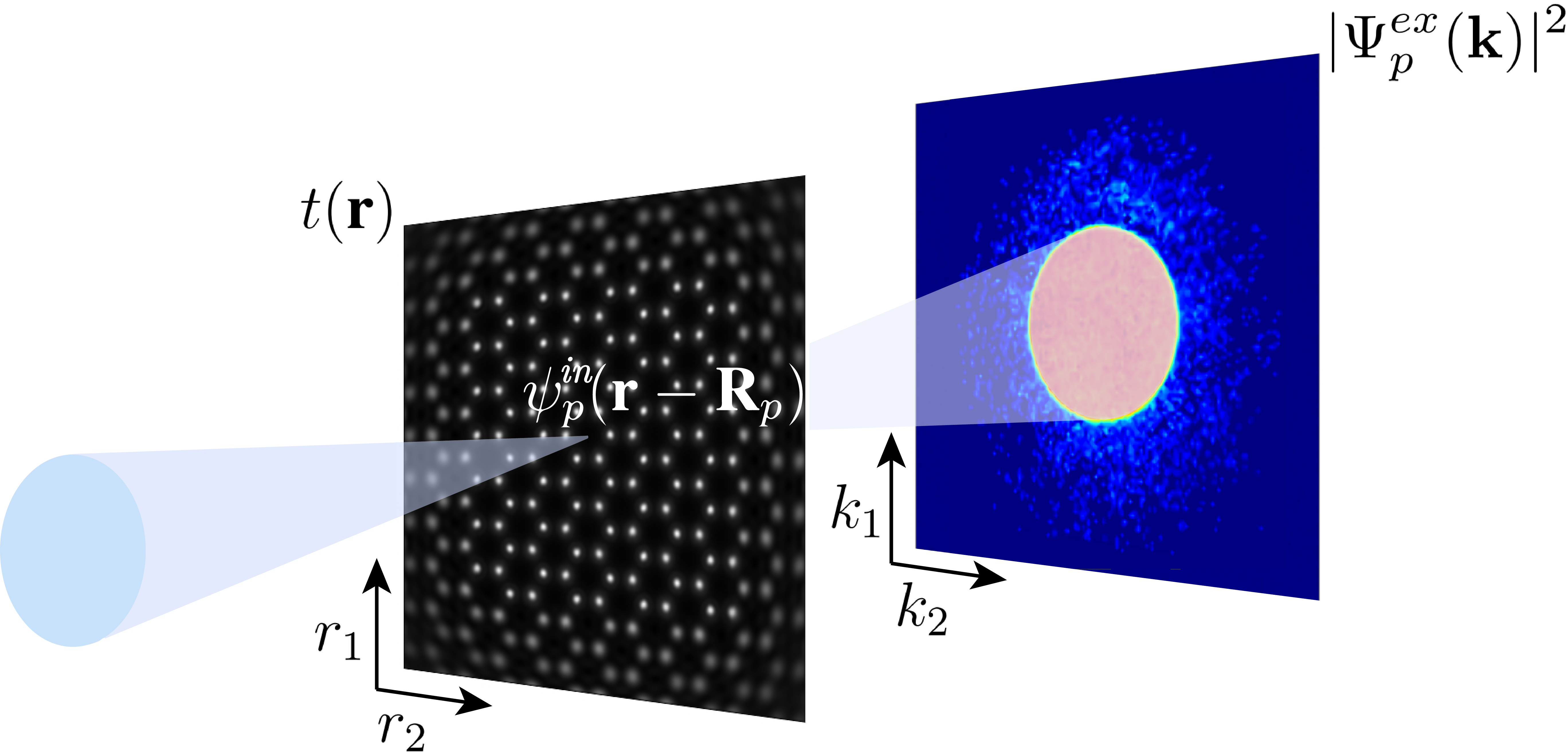}
\centering
\caption{Experimental setup in ptychography. At the scan position $\textbf{R}_p$ of the scan sequence, the beam illuminates a sample, where the incident electron wave $\psi^{in}_p(\textbf{r}-\textbf{R}_p)$ interacts with the transmission function $t(\textbf{r})$. The wave exiting the sample is propagated by a Fourier transform to the detector located in the far field and the intensity $I_p = |\Psi^{ex}_p(\textbf{k})|^2$ is recorded.}
\label{fig:scheme}
\end{figure}

Single slice ptychography can be expressed by a multiplicative approximation that describes the interaction of a wavefunction $\psi^{in}_p(\textbf{r})$ of an incoming beam with the transmission function $t(\textbf{r})$ of a specimen.
For each measurement $p$, the beam is shifted by $\textbf{R}_p$ and a diffraction pattern is acquired with the intensity $I_p$ that is expressed by: 
\begin{equation}
\label{eq:intensity}
\begin{split}
I_p = |\Psi^{ex}_p(\textbf{k})|^2  = |\mathcal{F} \left[ \psi^{in}_p(\textbf{r}-\textbf{R}_p)  t(\textbf{r}) \right]|^2,
\end{split}
\end{equation}
where $\mathcal{F}$ is the Fourier propagator, $\textbf{r}$ the real space coordinate, $\textbf{k}$ the reciprocal space coordinate and $\Psi^{ex}_p(\textbf{k})$ the exit wavefunction at the detector. According to the strong phase object approximation, the transmission function can be defined as $t(\textbf{r}) = e^{i \sigma V(\textbf{r})}$, with the interaction constant $\sigma$ and the complex quantity $V(\textbf{r})$, where the real part $V_{re}(\textbf{r})$ corresponds to the local projected electrostatic potential and the imaginary part $V_{im}(\textbf{r})$ accounts for absorption or scattering outside the range of scattering angles and energy losses recorded by the detector. Throughout the remainder of this paper, the variable $\sigma$ is absorbed into $V(\textbf{r})$.  X-ray and optical ptychography is mathematically described similarly with the only difference that the transmission function $t(\textbf{r})$ is related to the complex refractive index of the specimen. Figure \ref{fig:scheme} illustrates the experimental configuration of conventional ptychography.

The potential of the specimen is recovered from data of experimentally acquired diffraction patterns $J_p$ using a reconstruction algorithm. Here, we apply a gradient based algorithm \cite{schloz2020overcoming} with a gradient decent optimization and the potential is retrieved by iteratively minimizing the loss function:
\begin{equation}
\label{eq:ptycho}
\mathcal{L}(V) = \frac{1}{P}\sum^P_{p=1} \Vert I_p(V) - J_p\Vert^2_2 .
\end{equation}
Although the approach described in this paper is compatible with multisclice ptychography, in light of the application to a 2D material we constrain ourselves to single-slice ptychography.

%=========================================================

\subsection{Generation of scan sequences}
\label{sec:rnn}
\noindent We consider a recurrent neural network (RNN) \cite{rumelhart1986learning, elman1990finding, werbos1988generalization} for the generation of scan sequences. Its network architecture is designed to model temporal sequences with recurring input information. Memory cells combine the current input information $X_t$ with the hidden state $H_t$ and map it to the next hidden state $H_{t+1}$. These hidden states represent the memory gathered from all the previous time steps. At every time step $t$, an output is generated on the basis of the current hidden state. In the implementation shown here, the output corresponds to a sub-sequence of scan positions, given by a vector of 2D coordinates $\vec{\textbf{R}}_{P_t}$. In principle, the output can be reduced to a single scan position $\textbf{R}_{p_t}$, but we do not for to practical reasons that will be discussed later. The sub-sequence is predicted via a fully connected layer (FC) that is parameterized by the layer weights $\theta_H$:
\begin{equation}
\label{eq:H2R}
\vec{\textbf{R}}_{P_t} = \text{FC}_{\theta_H}(H_t).
\end{equation}

At the predicted scan positions $\vec{\textbf{R}}_{P_t}$, diffraction patterns $\boldsymbol{J}_{P_t}$ are acquired by the microscope and from these diffraction patterns a potential $V_t(\textbf{r})$ is reconstructed minimizing Eq. (\ref{eq:ptycho}). The intermediate reconstruction $V_t(\textbf{r})$ combined with its corresponding sub-sequence of scan positions $\vec{\textbf{R}}_{P_t}$ can then be used for the input information $X_t$ of the RNN. However, the bandwidth of the information given in $V_t(\textbf{r})$ and $\vec{\textbf{R}}_{P_t}$ differs strongly and thus pre-processing is required before the two components can be concatenated and mapped to $X_t$. For the processed location information $\textbf{L}_t$ based on the sub-sequence $\vec{\textbf{R}}_{P_t}$, a FC that is parameterized by the weights $\theta_R$ is used:
\begin{equation}
\label{eq:R2L}
\textbf{L}_t = \text{FC}_{\theta_R}(\vec{\textbf{R}}_{P_t}).
\end{equation}
For the processed structure information $\textbf{C}_t$ based on the reconstructed potential $V_t(\textbf{r})$, a compressed representation $z_t$ is generated by using the encoder part of a convolutional autoencoder \cite{masci2011stacked}. This processing step is described in more detail in appendix \ref{sec:auto}. The compressed representation $z_t$ is then fed into a FC that is parameterized by the weights $\theta_z$:
\begin{equation}
\label{eq:V2z}
\textbf{C}_t = \text{FC}_{\theta_z}(z_t).
\end{equation}
The processed location information $\textbf{L}_t$ is subsequently concatenated with the processed structure information $\textbf{C}_t$ and mapped to the input information $X_t$ with a FC that is parameterized by the weights $\theta_{LC}$. The whole process of predicting sub-sequences of scan positions and acquiring the corresponding diffraction patterns is repeated until a ptychographic dataset of desired size is reached.

\begin{figure}%[H]
\includegraphics[width=1.0\linewidth]{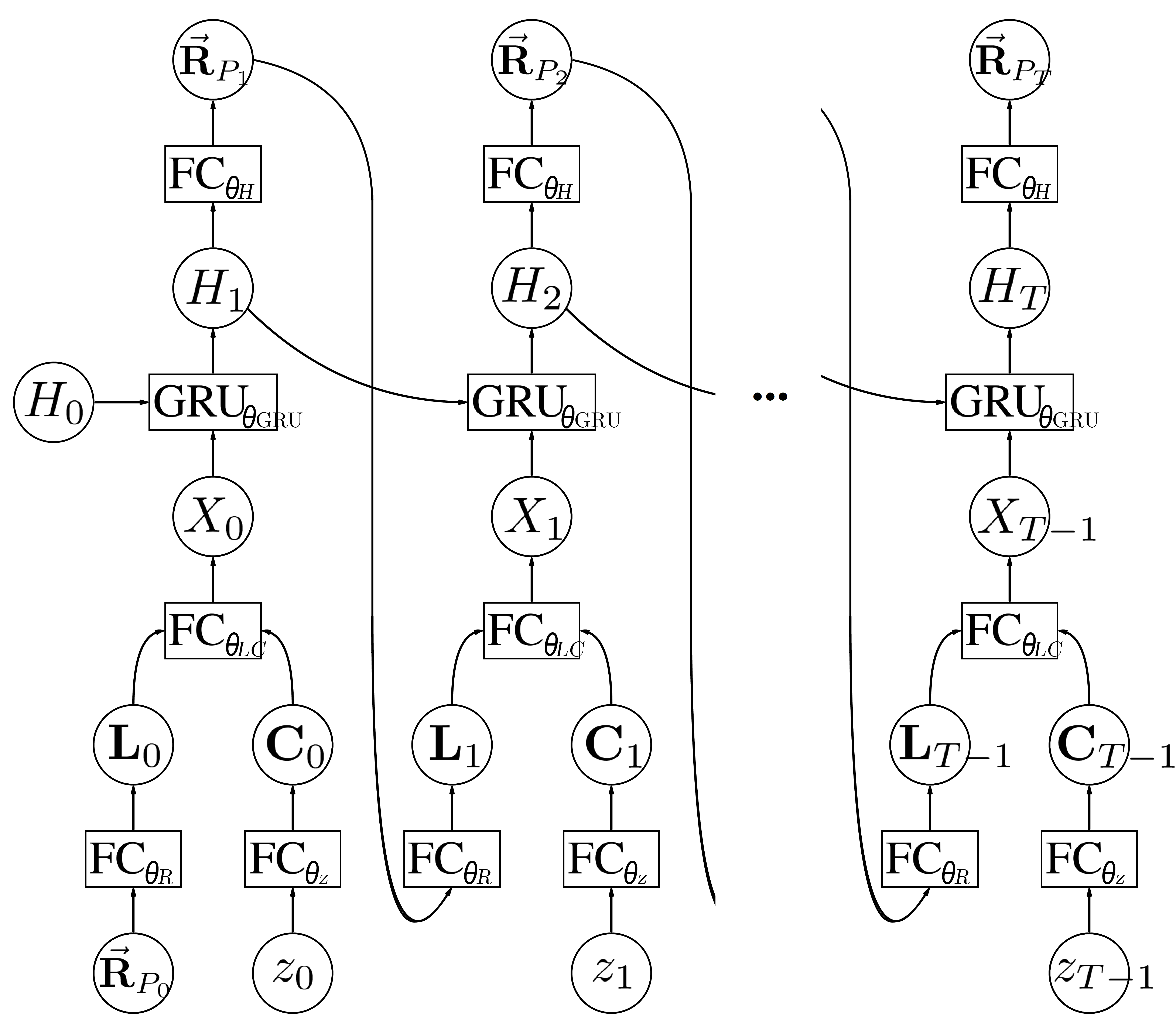}
\centering
\caption{Schematic of the forward propagation process of the RNN model. The RNN consists of GRU units that use the hidden state $H_t$ from the previous time step and the hybrid input information $X_t$ to create a new hidden state $H_{t+1}$. The hybrid input is the concatenation of the pre-processed information from the sub-sequence of scan positions $\vec{\textbf{R}}_{P_t}$ and the corresponding compressed representation of the partial reconstruction $z_t$. The output of the GRU cell is used to predict the positions of the next sub-sequence $\vec{\textbf{R}}_{P_{t+1}}$ and is also used as the input for the next GRU cell. The process is repeated until the full length of the scan sequence, consisting of $T$ sub-sequences, is reached.}
\label{fig:network}
\end{figure}

In practice and even after a reduction through adaptive scanning, several hundreds to thousands of diffraction patterns are required for effective ptychography. Covering this range of scan positions with a strong prediction performance requires efficient training of a large RNN. Backpropagation through time (BPTT) is typically used to generate the required gradients to update the network weights $\theta = \{ \theta_H,\theta_{\text{GRU}},\theta_{LC},\theta_R, \theta_z \}$ of the RNN. Its foundation on the chain-rule, with terms being multiplied by themselves as many times as the length of the network, can result in problems with training efficiency. For even the most basic RNN architectures, BPTT fails for relatively short sequences due to the so-called $\textit{vanishing}$ or $\textit{exploding}$ gradient problem \cite{bengio1994learning}. To circumvent this issue a more complex RNN architecture was proposed by Hochreiter et al. \cite{hochreiter1997long}. The Long-Short-Term-Memory (LSTM) network uses a more complex mapping between the input information and hidden state to the output, which allows a more efficient training using the BPTT method for larger networks. The gated recurrent units (GRU) network, which is a computationally faster, simplified version of the LSTM network, is used in this paper \cite{chung2014empirical}. A very large network would, nevertheless, be difficult to train using BPTT and also greatly increase acquisition time in adaptive scanning due to, e.g., a more frequent data transfer and generation of intermediate reconstructions $V_t(\textbf{r})$. Therefore, a sub-sequence of scan positions $\vec{\textbf{R}}_{P_t}$ is preferred over a single scan position $\textbf{R}_{p_t}$ as the RNN output. Figure \ref{fig:network} shows the prediction process modeled by the RNN in full detail.

\subsection{Training through reinforcement learning}
\label{sec:rl}
A RNN, such as the one described in the previous section, can be combined with RL to provide a formalism for modelling behaviour to solve decision making problems. In RL, a learning agent interacts with an environment, while trying to maximize a reward signal. This is generally formalized as a Markov decision process (MDP) described by a 5-tuple: $ \langle \mathcal{S},\mathcal{A},\rho,r,\gamma \rangle $. At each time-step $t$ the agent has complete knowledge of the environment by observing the state $s_t \in \mathcal{S}$ and makes an optimal decision by selecting an action $ a_t \in \mathcal{A}$. Based on $s_t$ and $a_t$, the next state $s_{t+1}$ is generated according to a transition function $\rho: \mathcal{S} \times \mathcal{A} \times  \mathcal{S} \rightarrow [0,1]$. The agent additionally receives a feedback through a scalar reward function $r: \mathcal{S} \times \mathcal{A} \rightarrow \mathbb{R}$. This reward $r$ contributes to the total reward computed at the end of the sequence, $G=\sum^T_{t=0} \gamma^{t} r(a_t,s_t) $, also known as the return. The discount factor $\gamma \in [0,1]$ controls the emphasis of long-term rewards versus short-term rewards.

In the case of adaptive scanning in ptychography, complete knowledge of the specimen structure is not known and the previous described formalism, where observations are equivalent to states, is not quite applicable. A partially observable Markov decision process (POMDP) generalizes the MDP to a 7-tuple: $ \langle \mathcal{S},\mathcal{A},\rho,r, \mathcal{O}, \omega, \gamma \rangle$ by considering the observation $o_t \in \mathcal{O}$ to contain only partial or incomplete information about the state $s_t$, and which is generated according to a observation function $\omega: \mathcal{A} \times \mathcal{S} \times \mathcal{O} \rightarrow [0,1]$. Therefore, $o_t$ can not sufficiently represent the state $s_t$ and instead the entire history of observations and actions up to the current time $h_t= \{ o_1, a_1, ..., o_{t-1},a_{t-1},o_t \}$ is used as basis for optimal or near-optimal decision making. A stochastic policy $\pi_{\theta}(a_t|h_t)$ maps the history of past interactions $h_t$ to action probabilities. Given a continuous action space, the policy can be represented by a two-dimensional Gaussian probability distribution:
\begin{equation}
\label{eq:policy}
\pi_{\theta}(a_t|h_t) = \mathcal{N}(\boldsymbol{\mu}_{\theta}(h_t), \Sigma),
\end{equation}
with its mean vector $\boldsymbol{\mu}_{\theta}(h_t)$ corresponding to $\textbf{R}_{p_t}$, where the history $h_t$ is summarized in the hidden state $H_t$ of the RNN and the covariance matrix $\Sigma$ with fixed variances $\sigma_x^2 \in [0,1]$ and $\sigma_y^2 \in [0,1]$. 

In this POMDP formalism, however, a single action $a_t$ is drawn from the probability distribution $\pi_{\theta}(a_t|h_t)$, which corresponds to a single agent interacting with the environment. This is incompatible with scan control in ptychography where we seek to predict multiple scan positions at each time step. A partially observable stochastic game (POSG) extends the POMDP formalism to a 8-tuple, $ \langle M, \mathcal{S},\{\mathcal{A}^m\}_{m\in M},\rho,\{r^m\}_{m\in M}, \{\mathcal{O}^m\}_{m\in M}, \omega, \gamma \rangle$, with multiple agents $M$, each selecting an action $a^m_t$ and making an observation $o^m_t$ given the state $s_t$. Thus, joint actions $\boldsymbol{a}_t = \langle a^1_t, ..., a^m_t \rangle$ from the joint action space $\boldsymbol{\mathcal{A}} = \mathcal{A}_1 \times ... \times \mathcal{A}_M $ are executed and joint observations $\boldsymbol{o}_t = \langle o^1_t, ..., o^m_t \rangle$ from the joint observation space  $\boldsymbol{\mathcal{O}} = \mathcal{O}_1 \times ... \times \mathcal{O}_M $ are received from the environment at each time step. In this case, the transition function is given by  $\rho: \mathcal{S} \times \boldsymbol{\mathcal{A}} \times \mathcal{S} \rightarrow [0,1]$, the observation function is given by $\omega: \boldsymbol{\mathcal{A}} \times \mathcal{S} \times \boldsymbol{\mathcal{O}} \rightarrow [0,1]$ and each agent receives its immediate reward defined by the reward function $r^m: \mathcal{S} \times \boldsymbol{\mathcal{A}} \rightarrow \mathbb{R}$. Here, we consider the individual agent to have access to the actions and observations of all other agents, which allows the optimization of its individual policy $\pi_{\theta^m}(a^m_t|\boldsymbol{h}_t)$ using the joint history of observations and actions $\boldsymbol{h}_t= \{ \boldsymbol{o}_1, \boldsymbol{a}_1, ..., \boldsymbol{o}_{t-1},\boldsymbol{a}_{t-1},\boldsymbol{o}_t \}$. The joint policy of all agents is then defined as $\boldsymbol{\pi}_{\boldsymbol{\theta}}(\boldsymbol{a}_t| \boldsymbol{h}_t) = \prod_{m=1}^{M} \pi_{\theta^m}(a^m_t|\boldsymbol{h}_t)$, with $\boldsymbol{\theta}=\{ \theta^m \}_{m \in M}$. The goal of RL is now to learn a joint policy that maximizes the expected total reward for each agent $m$ with respect to its parameters $\theta^m$: 
\begin{equation}
\label{eq:objectiveRL}
\begin{split}
\mathcal{J}^m(\boldsymbol{\theta})  =  \mathbb{E}_{\boldsymbol{\pi}_{\boldsymbol{\theta}}(\boldsymbol{\tau})}  \left[ G^m \right] \approx \frac{1}{N} \sum^N_{n=1} G^m_n,
\end{split}
\end{equation}
where the expected total reward can be approximated by Monte Carlo sampling with $N$ samples. In this paper, improvement of the policy is achieved by updating the policy parameters $\theta^m = \{ \theta^m_H,\theta_{\text{GRU}},\theta_{LC},\theta_R, \theta_z \} $ with 'REINFORCE' \cite{williams1992simple}, a policy gradient method:
\begin{equation}
\label{eq:gradientRL}
\begin{split}
\nabla_{\theta^m} \mathcal{J}^m(\boldsymbol{\theta}) = \mathbb{E}_{\boldsymbol{\pi}_{\boldsymbol{\theta}}(\boldsymbol{\tau})}  \left[ \nabla_{\theta^m} \text{log} \boldsymbol{\pi}_{\boldsymbol{\theta}}(\boldsymbol{\tau})  G^m \right]  \\ \approx \frac{1}{N} \sum^N_{n=1} \sum^T_{t=0} \nabla_{\theta^m} \text{log} \pi_{\theta^m}(a^m_{n,t}|\boldsymbol{h}_{n,t}) & G^m_n.
\end{split}
\end{equation}
The derivation of $\nabla_{\theta^m} \mathcal{J}^m(\boldsymbol{\theta})$ is given in the appendix \ref{app:REINFORCE}.

\subsection{Learning to adaptively scan in ptychography} 
While policy gradient methods are the preferred choice to solve reinforcement learning problems in which the action spaces are continuous \cite{liu2020improved}, they come with significant problems. Like any gradient based method, policy gradient solutions mainly converge to local, not global, optima \cite{sutton2000policy}. In this paper, we reduce the effect of this problem during training by splitting the training of the RNN into supervised learning and RL. The first training step initializes the policy parameters such that the scan pattern follows a conventional grid pattern, thereby avoiding relatively poor local optima during subsequent policy gradient steps. This training step is explained in more detail in the appendix \ref{sec:policyinitialization}. In the second training step, the pre-trained policy is fine tuned through RL, resulting in a scan pattern that has been adapted to the structure of the material.

A high variance of gradient estimates is another problem that particularly strongly affects the Monte Carlo policy gradient method \cite{wu2018variance, liu2020improved, feriani2021single}. Due to this, the sampling efficiency is relatively low, which causes a slow convergence to a solution. This makes deep RL applied to ptychography challenging as the image reconstruction itself requires iterative processing (see section \ref{sec:ptycho}). 

The high variance can be in part attributed to the difficulty of assigning credit from the overall performance to an individual agent's action. This credit assignment problem is limited to a temporal problem in the single agent RL case \cite{agogino2004unifying}. In this case, methods for variance reduction to better assign credit to individual actions are for instance reward-to-go \cite{di2012policy} or using a baseline \cite{sutton2000policy, wu2018variance}. 

In RL involving multiple cooperating agents with a shared reward function $r^1(\boldsymbol{a}_t|s_t)=r^2(\boldsymbol{a}_t|s_t)=...=r^m(\boldsymbol{a}_t|s_t)$, the challenge to overcome the credit assignment problem further increases due to the necessity of now identifying the contribution of each agent's action to the total reward. This challenge can be tackled with difference reward \cite{wolpert2002optimal, tumer2007distributed, colby2014approximating, castellini2020difference}, which replaces the shared reward with a shaped reward that is formed by comparing the global reward with a reward that an agent would receive when performing a default action.

\begin{figure}%[H]
\includegraphics[width=1.00\linewidth]{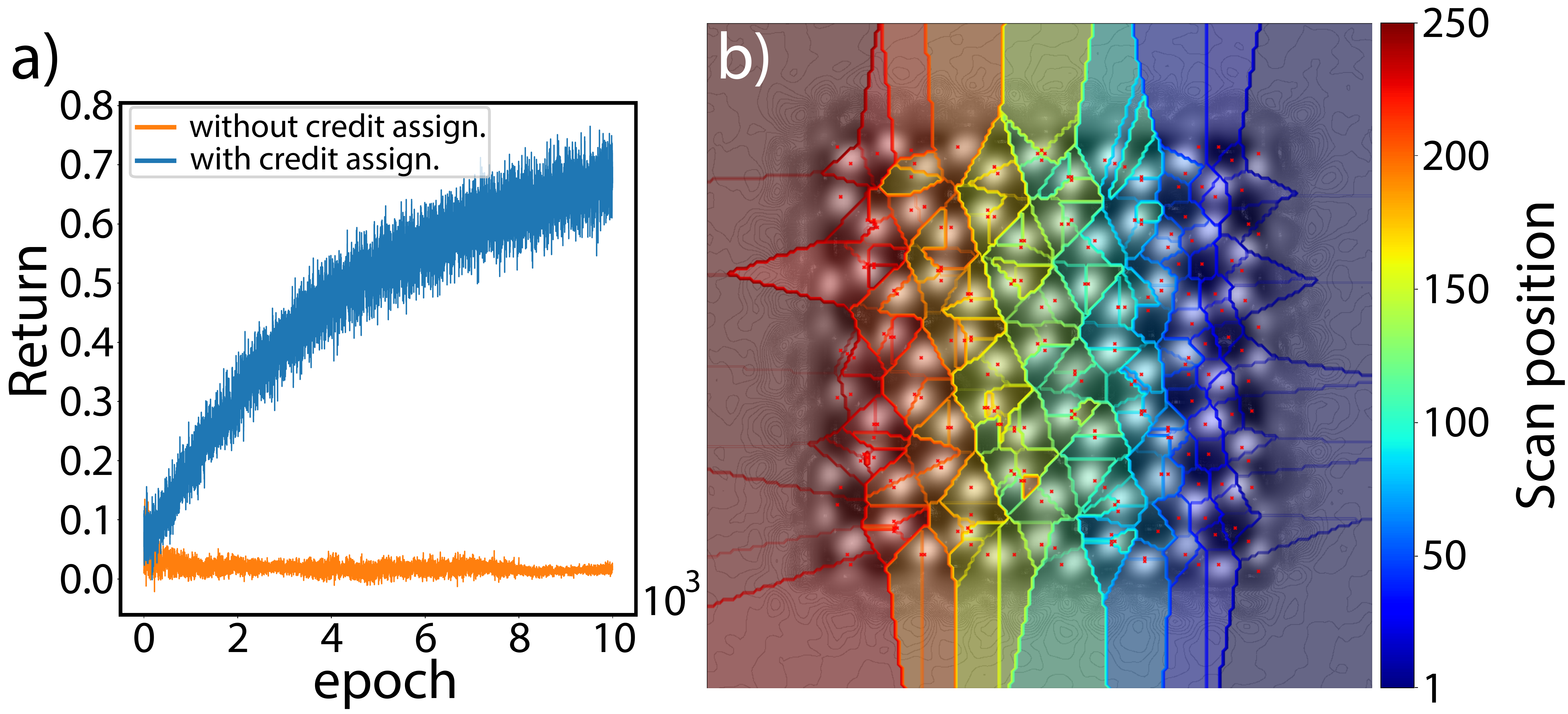}
\centering
\caption{a)  Learning curves of RL with multiple agents that use a shared reward or shaped reward, illustrated in orange and blue, respectively. b) A Voronoi diagram is used to assign a unique reward to each scan position of the predicted sequence. The scan positions are shown as red dots, where the first 50 positions are distributed on the right side within the dark blue area. For visualization purpose, the ground truth reconstruction is included in the diagram.}
 \label{fig:voronoi}
\end{figure}

Following the idea of a difference reward in spirit, we introduce a way to estimate the reward function in order to tackle the credit assignment problem for adaptive scanning in ptychography. The reward function should naturally correspond to the quality of the ptychographic reconstruction. We have found empirically that a high reconstruction quality correlates positively with a high dynamic range in the phase. Therefore, the reward function could intuitively be formalized by $r^m(\boldsymbol{a}_t|s_t)=P^{-1} \sum_{\textbf{r}\in \text{FOV}} V(\textbf{r})$, where $P$ is the total number of scan positions. This formulation, however, does not solve the credit assignment problem and results in an insufficient training performance, as shown in Figure \ref{fig:voronoi}a). To estimate the reward for the actions of each individual agent, we use a tessellation method that partitions the atomic potential into small segments. A Voronoi diagram \cite{okabe2016spatial}, where each position corresponds to a seed for one Voronoi cell, enables assignment of only a part of the total phase to each position. More precisely, the Voronoi diagram formed by the predicted scan positions is overlaid with the corresponding ptychographic reconstruction at the end of the prediction process and the summed phase within each Voronoi cell is the reward for that cell's seed position. The reward function can be expressed by $r^m(\boldsymbol{a}_t|s_t)= P^{-1} \sum_{\textbf{r}\in \text{Cell}^m} V(\textbf{r}) $. Figure \ref{fig:voronoi}b) shows a Voronoi diagram generated by predicted scan positions. 

\subsection{Experiment and model design}
\label{seq:settings}
For the experimental investigation, we acquired multiple ptychographic datasets from a monolayer molybdenum disulfide (MoS$_2$) specimen with a NION HERMES microscope. The microscope was operated with a $60$~kV acceleration voltage, a convergence angle of $33$~mrad and diffraction patterns with a pixel size of $0.84$~mrad were acquired using a Dectris ELA direct electron detector mounted at the electron energy loss spectroscopy (EELS) camera port. Distortions induced by the EEL spectrometer were corrected with in-house developed software. For the ptychographic dataset acquisition, a conventional grid scan with a scanning step size of $0.02$~nm was used. 

From the experimentally acquired datasets we created $175$ smaller datasets, each with 10,000 diffraction patterns. The diffraction patterns were binned by a factor of $2$ to $64 \times 64$ pixels. The adaptive scanning algorithm was then trained on the smaller datasets with the goal of predicting optimal scan sequences of $250$ to $500$ probe positions, out of the possible 10,000, which corresponds to a dose reduction by a factor of $40$ to $20$. Each sub-sequence contains 50 to 100 positions, where the first sub-sequence follows a quasi-random Halton sequence. 

The ptychographic reconstructions were performed with an optimized version of ROP \cite{schloz2020overcoming} that allows simultaneous reconstruction from a batch of different datasets, which was required for efficient model training. A gradient descent step size $\alpha_{\text{ROP}}$ of $5.25\ensuremath{\text{\textsc{e}}}2$ was chosen and the potential was retrieved at iteration 5. The reconstructed potential was $200\times 200$ pixels with a pixel size of $0.0154$~nm, for a field of view of $2 \times 2$~nm. For the generation of the reward function, Voronoi diagrams were generated with the Jump Flooding Algorithm \cite{rong2006jump} and for the implementation of the network models, PyTorch \cite{NEURIPS2019_9015} was used. For the compression of structure information, we used a convolutional autoencoder consisting of $6$ convolutional layers with kernels of dimension $3$, a stride of $1$ and channels that ranged from $16$ to $512$ for the encoder and decoder part, respectively. The input of the autoencoder had a dimension of $512$ with a pixel size of $0.0064$~nm and thus a scaling and an interpolation was required before the potential generated by ROP could be compressed. In addition, the value of the potential $V_i$ at each pixel $i$ was transformed to zero mean and unit variance. For the prediction of the scan sequences, pre-training and fine-tuning was performed with a RNN model composed of $2$ stacked GRU layers with hidden states $H_t$ of size $2048$, the Adam optimizer \cite{kingma2014adam} with a learning rate $\alpha_{\text{RNN}}$ of $1\ensuremath{\text{\textsc{e}-}}5$ and a batch size of $24$. For the fine-tuning, a policy with variances of $\sigma_x^2 = \sigma_y^2 = 0.0125^2$ was chosen and a myopic behavior was enforced by setting the discount factor for the return, $G$, to $\gamma = 0$. All settings used for training the adaptive scanning algorithm are summarized in Table \ref{tab:settings}. 

\section{Results}
\subsection{Adaptive scanning on experimental MoS$_2$ data}
\begin{figure}
\includegraphics[width=1.0\linewidth]{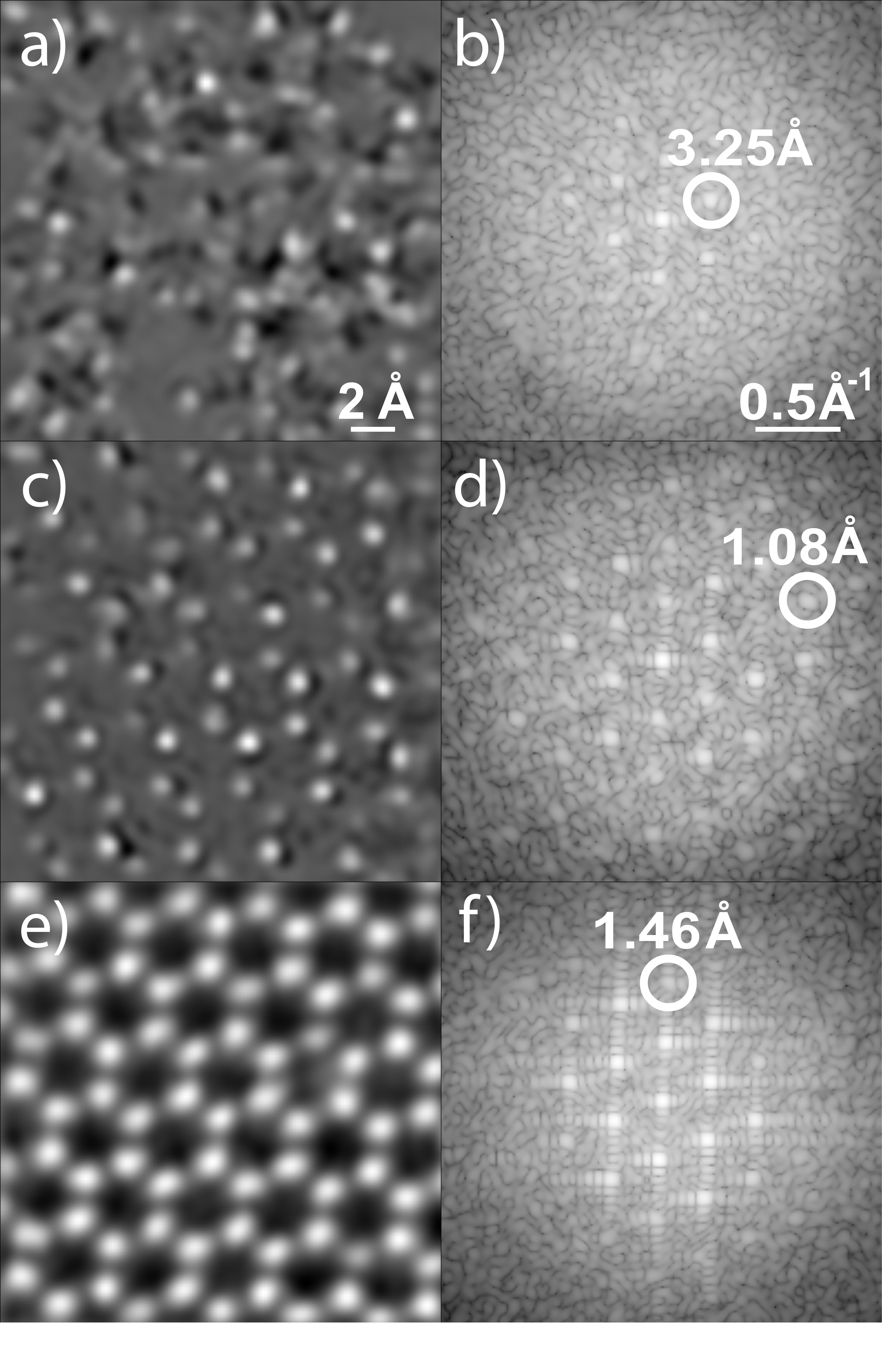}
\centering
\caption{Ptychographic reconstructions of a MoS$_2$ data set with different scanning procedures. a) Reconstruction from $250$ diffraction patterns of the data set that correspond to scan positions which follow a random sequence and c) an adaptively predicted sequence. e) Reconstruction of the full data set with 10,000 diffraction patterns acquired with the conventional grid scan. b), d) and f) Corresponding diffractograms of the reconstructions. The real-space distance of the circled diffraction spots is labeled.}
\label{fig:result}
\end{figure}

Figure \ref{fig:result} shows the result of adaptive scanning on experimentally acquired MoS$_2$ data and compares it to the result of a random scanning and the conventional grid scanning procedure. The data used for the comparison was not part of the training data for the adaptive scanning model. While the full data set consisting of 10,000 diffraction patterns has been used to obtain a ground truth reconstruction, only $250$ diffraction patterns have been used for the adaptive scanning as well as the random scanning reconstruction. Figure \ref{fig:result}a) shows the ptychographic reconstruction when using a random scanning procedure. The structure of the material is not clearly resolved and large parts of the field of view are not covered by the scanning procedure. Figure \ref{fig:result}c) shows the reconstruction when the scan positions are predicted by the adaptive scanning algorithm. The structure of the MoS$_2$ material is now much better resolved and is closer to the ground truth reconstruction of the full data grid scanning procedure, shown in Figure \ref{fig:result}e). Figure \ref{fig:result}b), d) and f) show the diffractograms of the corresponding reconstructions and the circled diffraction spots show that the highest resolution of $1.08$~\AA{}  is achieved by the adaptive scanning procedure, while the lowest resolution of $3.25$~\AA{} is obtained by the random scanning procedure. Further examples of reconstructions and their corresponding scan sequences are shown in Figure \ref{fig:result_examples}. 

The results suggest that probe delocalization due to scattering plays an important role as to why an improved ptychographic reconstruction can be achieved by distributing the scan positions predominantly on the atoms of the specimen. When the beam is positioned on an atom, it scatters to higher angles and thus experiences spatial delocalization. It therefore also probes a larger range in real space, i.e. the scattering includes the local environment of the atom the beam hits. This implies that similar results could be achieved by using RL with a reward function that specifically emphasizes the scattered electrons in the recorded diffraction patterns, which is an interesting area for future research.

The final point of our investigation into adaptive scanning in ptychography evaluates the performance of the method for various prediction settings. We compare the structural similarity index measure (SSIM) \cite{wang2004image} between the reconstruction obtained from the reduced data and the ground truth reconstruction obtained from the full data to quantify the improvement when using adaptive scanning. Here, SSIM$_a$ and SSIM$_r$ are the SSIM  using a reconstruction of reduced data that is obtained with the adaptive scanning and the random scanning procedure, respectively. Table \ref{tab:Realizations} shows the relative reconstruction quality improvement $Q_{\text{SSIM}} = ( \text{SSIM}_a - \text{SSIM}_r ) / (\text{SSIM}_r)$ for different experimental settings averaged over $25$ data sets. Additionally, the relative resolution improvement $Q_{\text{res}}$ averaged over the same datasets is given. In the case of $250$ scan positions, which corresponds to a dose reduction by a factor of $40$ with respect to the original data, tests were performed for multiple sub-sequences, i.e. predictions. The quality improvement ranges from $16.71\%$ to $25.75\%$ and the resolution improvement ranges from $9.74\%$ to $27.57\%$ for a number of $2$ to $5$ sub-sequences (corresponding to 1 to 4 predictions), respectively. Further tests were performed using a larger number of total scan positions and 5 sub-sequences. However, while the relative resolution slightly improves for an increasing amount of scan positions, the difference in quality between the reconstruction generated with the positions of the adaptive scan and the random scan decreases with the total number of positions used, as can be expected, since the random sampling covers the sampled area in an increasingly complete manner. It should be noted that for all tests which used adaptive scanning with $5$ sub-sequences, a higher resolution of the reconstruction compared to that of the reconstruction of the full data set was achieved. These results indicate that the reconstruction quality and resolution improves with the frequency by which the positions are predicted, and that low dose experiments benefit the most from the adaptive scanning scheme. 

\begin{table}
\centering
\caption{Performance of adaptive scanning for various experimental settings that differ in the number of scan positions and the total number of sub-sequences. For each setting, the oversampling ratio $N_\text{k}/N_\text{u}$, which is calculated following \cite{schloz2020overcoming}, and the electron dose is given.}
\label{tab:Realizations}
\begin{tabular}{ c c c c | c | c }
\hline
\hline 
\# Pos. & $N_\text{k}/N_\text{u}$ 	  	& Dose  ($\mathrm{e^-}$/\AA$^{-2}$)		& \# Sub-seq.  	& Q$_{\text{SSIM}}$ & Q$_{\text{res}}$  \\
  \hline
 $250$ 			& $8.21$  & $1.34 \text{\ensuremath{\text{\textsc{e}}}5 }$  & $2$ 	& 16.71\% &$9.74\%$\\ %\pm $11.12\%$
               &    & 	& $3$ & 22.89\% &$15.06\%$\\ %\pm $18.30\%$
               &    & 	& $4$ & 24.96\% &$19.79\%$ \\ %\pm $12.62\%$
               &    & 	& $5$ & $25.75\%$ & $27.57\%$\\ %\pm $12.24\%$
  \hline
 $335$ 			& $10.83$  & $1.79 \text{\ensuremath{\text{\textsc{e}}}5 }$   & $5$ 	&  $12.53\%$ & $28.71\% $	\\ %\pm $12.82\%$
  \hline 
 $420$ 			& $13.43$  & $2.25 \text{\ensuremath{\text{\textsc{e}}}5 }$   & $5$ 	&  $12.29\% $ &	$29.23\% $\\ %\pm $11.17\%$
  \hline 
 $500$ 			& $15.86$  & $2.68 \text{\ensuremath{\text{\textsc{e}}}5 }$   & $5$ 	&  $8.16\% $ & $31.59\% $	\\ %\pm $11.58\%$
  \hline 
  \hline
\end{tabular}
\end{table}

\section{Conclusion}

In this paper we present a method for electron ptychography that reduces the electron dose through adaptive scanning. It is based on the idea that ptychography requires only an increased information redundancy through overlapping illuminating beams at regions of the sample that contain atomic structure. The prediction algorithm is a mixture of deep learning models being trained using supervised and reinforcement learning.

We show an improved reconstruction quality and resolution when using an adaptive scanning approach on experimentally acquired monolayer MoS$_2$ datasets in comparison to another dose reduction scanning approach. In a low dose experiment the adaptive scanning procedure could improve on average the reconstruction quality by up to $27.75\%$ and the resolution by up to $31.59\%$. The resolution achieved by adaptive scanning was also higher than that of the reconstruction from the full data set, but without the same homogeneous reconstruction quality throughout the entire field of view. %For a dose reduction by a factor of about $40$, $30$ and $24$, the resolution could be improved by $20\%$, $35\%$ and $42\%$, respectively.

These improvements show that adaptive scanning for ptychography is a useful technique to lower the dose needed for the analysis of sensitive samples. In addition to that, the proposed algorithm can be taken as a blueprint for a broad range of scanning based microscopy methods and thus paves the way for future research in machine learning supported, automated and autonomous microscopy.

\begin{acknowledgments}
\noindent M.S., T.C.P. and W.V.d.B. acknowledge financial support from the Deutsche Forschungsgemeinschaft (DFG, German Research Foundation) Grant No. BR 5095/2-1. M.S., J.M. and C.T.K acknowledge support from the DFG - Project-ID 182087777 - SFB 951. M.S. and C.T.K. acknowledge financial support from the DFG - Project-ID 414984028 - SFB 1404. We thank Prof. Sang Ho Oh and Dr. Jinsol Seo (Korea Institute of Energy Technology, Naju, Korea), as well as Dr. Bumsu Park (CEMES, Toulouse, France) for providing the MoS$_2$ sample. M.S. would like to thank Jayesh K. Gupta (Microsoft, Stanford, USA) for his helpful suggestion about the description of the POSG formalism.
\end{acknowledgments}

%\FloatBarrier

\appendix

\section{Structure content compression through convolutional autoencoders}
\label{sec:auto}
\begin{figure}[H]
\includegraphics[width=0.9\linewidth]{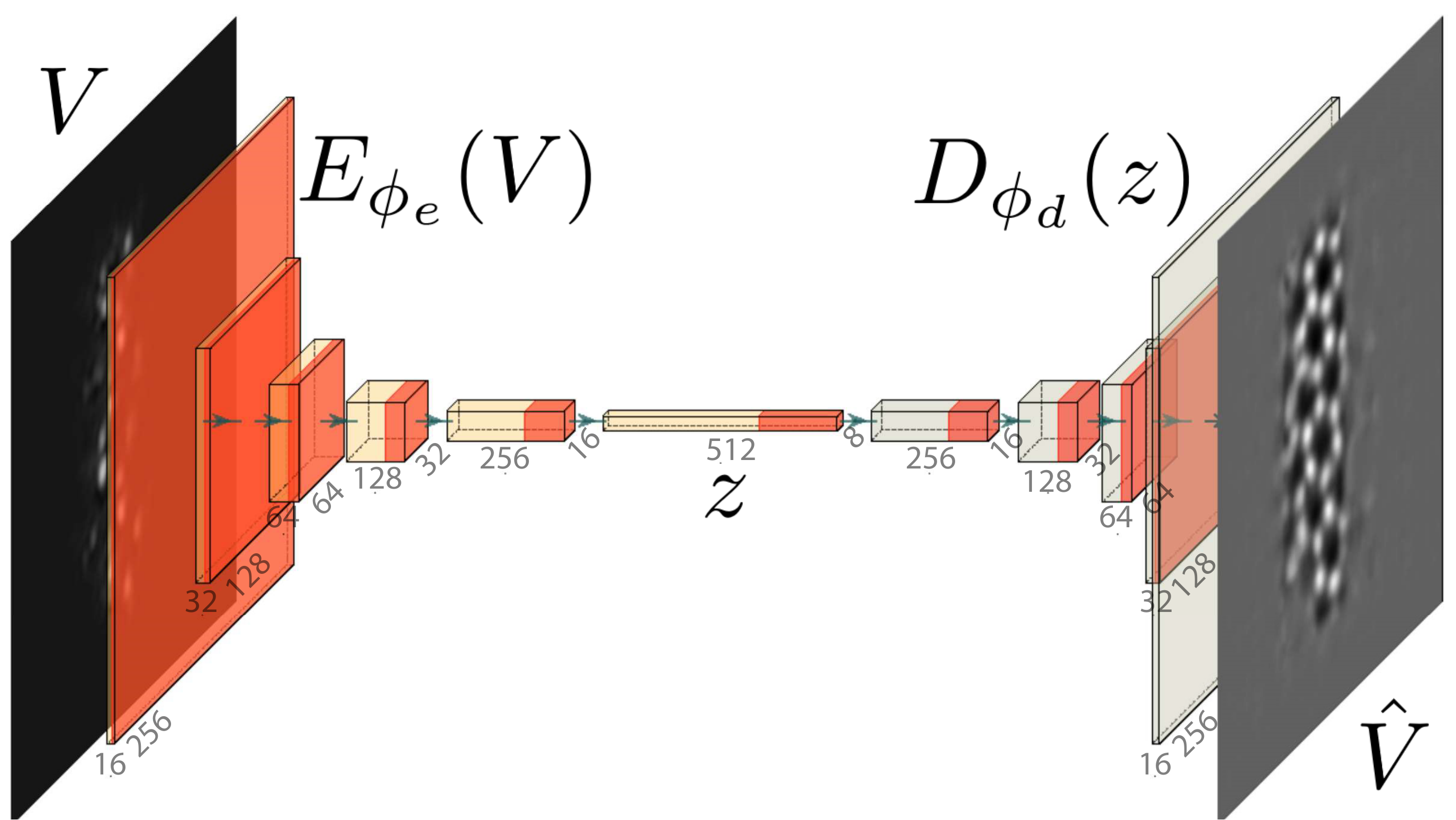}
\centering
\caption{Schematic of the convolutional autoencoder model. A reconstruction $V$ generated from diffraction patterns is mapped to the compressed representation $z$ by using the encoder network $E_{\phi_e}(V)$. The compressed representation $z$ is then basis for a reverse mapping by the decoder network $D_{\phi_d} ( z )$ to generate a prediction of the potential $\hat{V}$. }
\label{fig:auto}
\end{figure}

For the processing of the intermediate reconstruction $V_t(\textbf{r})$ that is the basis for the compressed representation $z_t$ of the RNN, we make use of a convolutional autoencoder \cite{masci2011stacked}, which is a deep learning model based on convolutional neural networks \cite{lecun1989backpropagation}. This model is composed of two parts, namely an encoder and a decoder. The encoder transforms potentially degraded input data into a compressed representation and the decoder recovers the original input from this representation \cite{hinton2006reducing}. By reducing the dimensionality, the model ensures that only the most important information for the reconstruction is extracted. In the convolutional autoencoder applied in this paper, an image $V \in \mathbb{R}^{H \times W \times 2}$ reconstructed from diffraction patterns is mapped to the latent representation $z \in \mathbb{R}^{\frac{H}{2^b} \times \frac{W}{2^b}\times f}$  by using the encoder network $E_{\phi_e}$:
\begin{equation}
    z = E_{\phi_e} ( V ),
\end{equation}
where $\phi_e$ corresponds to the encoder network weights. The encoder network consists of a concatenation of $b$ convolution layers that increase the feature space $f$ and halve the dimension size $H \times W$. The latent representation $z$ is then mapped back to the original dimensions by the decoder network:
\begin{equation}
    \hat{V} = D_{\phi_d} ( z ),
\end{equation}
where $\phi_d$ are the decoder network weights and $\hat{V}$ is the predicted image. Transposed convolutional layers form the decoder network, which has the inverse effect to the layers used in the encoder network. Figure \ref{fig:auto} illustrates the encoder-decoder architecture in full detail.

\begin{figure}[H]
\includegraphics[width=1.0\linewidth]{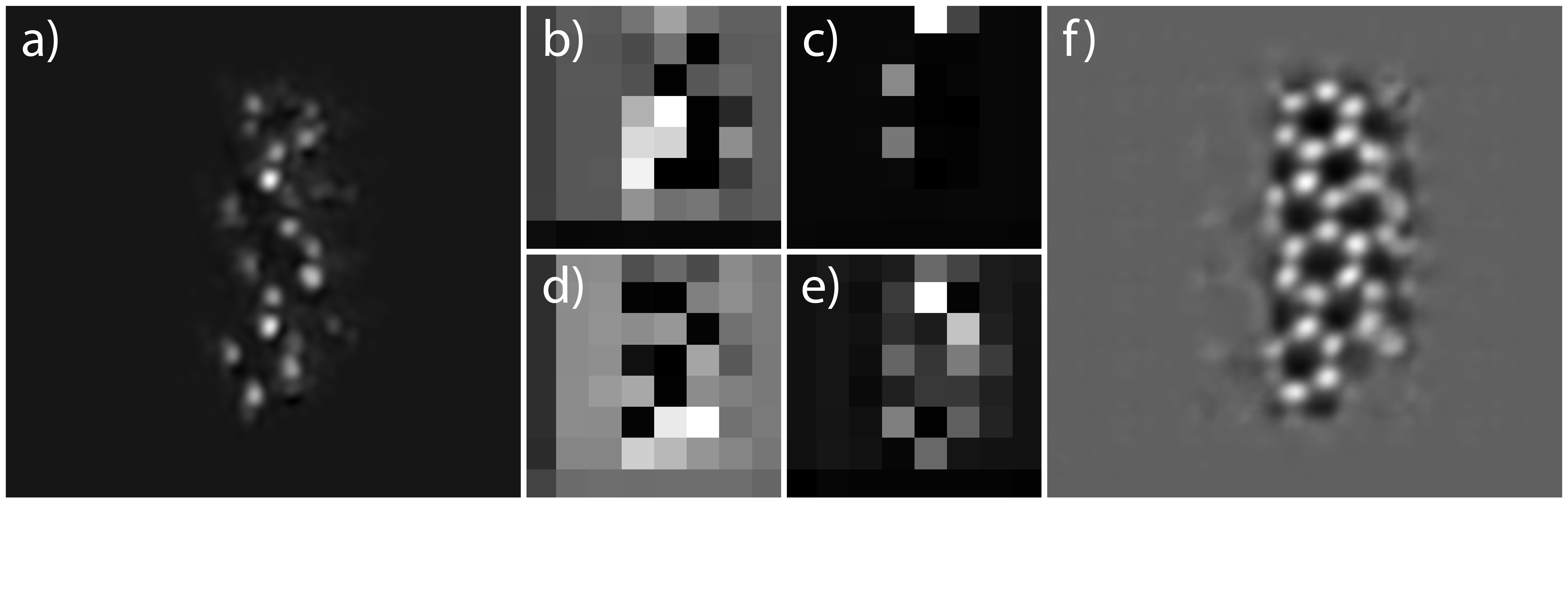}
\centering
\caption{Convolutional autoencoder applied on partial structure information, given by the reconstruction of data from a sub-sequence of scan positions. a) The reconstruction $V_t(\textbf{r})$ from a sub-sequence of scan positions that is used as input for the convolutional autoencoder. b)-e) $4$ channels of the compressed representation $z_t$ of the structure information. f) Decoded structure information $\hat{V}_t$ from $z_t$.}  
\label{fig:CompAuto}
\end{figure}

After estimating the network weights $\phi_e$ and $\phi_d$ by minimizing the loss function: 
\begin{equation}
\mathcal{M}(\phi_e, \phi_d) = \Vert  D_{\phi_d} ( E_{\phi_e}( V )) - V \Vert^2_2,
\end{equation}
we can utilize the encoder network $E_{\phi_e}$ for the compression of $V_t(\textbf{r})$. Figure \ref{fig:CompAuto} shows a compression of a partial reconstruction $V_t(\textbf{r})$ and the decompression of its corresponding compressed representation $z_t$. This pre-processing helps the algorithm form the hybrid input information $X_t$ by reducing the structure input information size, but also handle reconstructions from experimentally acquired data that may suffer from noise, contamination and/or incorrect scan positions.

\section{The "REINFORCE" algorithm}
\label{app:REINFORCE}
In the case of the multi agent RL problem, where we use the POSG formalism, the objective of an agent $m$ given by Eq. (\ref{eq:objectiveRL}) can be expressed by: 
\begin{equation}
\label{eq:costRLDerivation}
\mathcal{J}^m(\boldsymbol{\theta})  = \mathbb{E}_{\boldsymbol{\pi}_{\boldsymbol{\theta}}(\boldsymbol{\tau})}  \left[ G^m \right] = \int \boldsymbol{\pi}_{\boldsymbol{\theta}}(\boldsymbol{\tau})  G^m d\boldsymbol{\tau} ,
\end{equation}
with the trajectory $\boldsymbol{\tau} = \lbrace s_0, \boldsymbol{o}_0, \boldsymbol{a}_0, s_1, ..., s_T, \boldsymbol{o}_T, \boldsymbol{a}_T \rbrace$ and the policy induced trajectory distribution  $\boldsymbol{\pi}_{\boldsymbol{\theta}}(\boldsymbol{\tau}) = q(s_0) \prod^T_{t=0} \rho(s_{t+1}|s_t, \boldsymbol{a}_t) \boldsymbol{\pi}_{\boldsymbol{\theta}}(\boldsymbol{a}_t | \boldsymbol{h}_t) \omega(\boldsymbol{o}_t|s_t)$ and where $q(s_0)$ is the distribution of initial states. Applying the gradient $\nabla_{\theta^m}$ to the objective and using the identity $\nabla_{\boldsymbol{\theta}} \boldsymbol{\pi}_{\boldsymbol{\theta}}(\boldsymbol{\tau}) = \boldsymbol{\pi}_{\boldsymbol{\theta}}(\boldsymbol{\tau}) \nabla_{\boldsymbol{\theta}} \text{log} \boldsymbol{\pi}_{\boldsymbol{\theta}}(\boldsymbol{\tau})$, we obtain:
\begin{equation}
\label{eq:gradientREINFORCE}
\begin{split}
\MoveEqLeft [40]
\nabla_{\theta^m} \mathcal{J}^m(\boldsymbol{\theta})  = \int \boldsymbol{\pi}_{\boldsymbol{\theta}}(\boldsymbol{\tau})  \nabla_{\theta^m} \text{log} \boldsymbol{\pi}_{\boldsymbol{\theta}}(\boldsymbol{\tau}) G^m d\boldsymbol{\tau}
\cr
\MoveEqLeft [36.2]
= \int \boldsymbol{\pi}_{\boldsymbol{\theta}}(\boldsymbol{\tau})  \nabla_{\theta^m} \text{log} [q(s_0) \prod^T_{t=0} \rho(s_{t+1}|s_t, \boldsymbol{a}_t) 
\cr
\MoveEqLeft [33]
\times \prod^M_{m=1} \pi_{\theta^m}(a^m_t | \boldsymbol{h}_t) \omega(\boldsymbol{o}_t|s_t)]  G^m d\boldsymbol{\tau}
\cr
\MoveEqLeft [36.2]
=  \mathbb{E}_{\boldsymbol{\pi}_{\theta}(\boldsymbol{\tau})}  \Biggl[ \sum^T_{t=0} \nabla_{\theta^m} \text{log} \pi_{\theta^m}(a^m_t|\boldsymbol{h}_t)
\cr
\MoveEqLeft [33]
\times  \left( \sum^T_{t^{'}=t} \gamma^{t^{'}-t} r^m(\boldsymbol{a}_{t^{'}},s_{t^{'}})  \right) \Biggr]
\cr
\end{split}
\end{equation}

\section{Policy initialization through supervised learning}
\label{sec:policyinitialization}
\begin{figure}[H]
\includegraphics[width=1.00\linewidth]{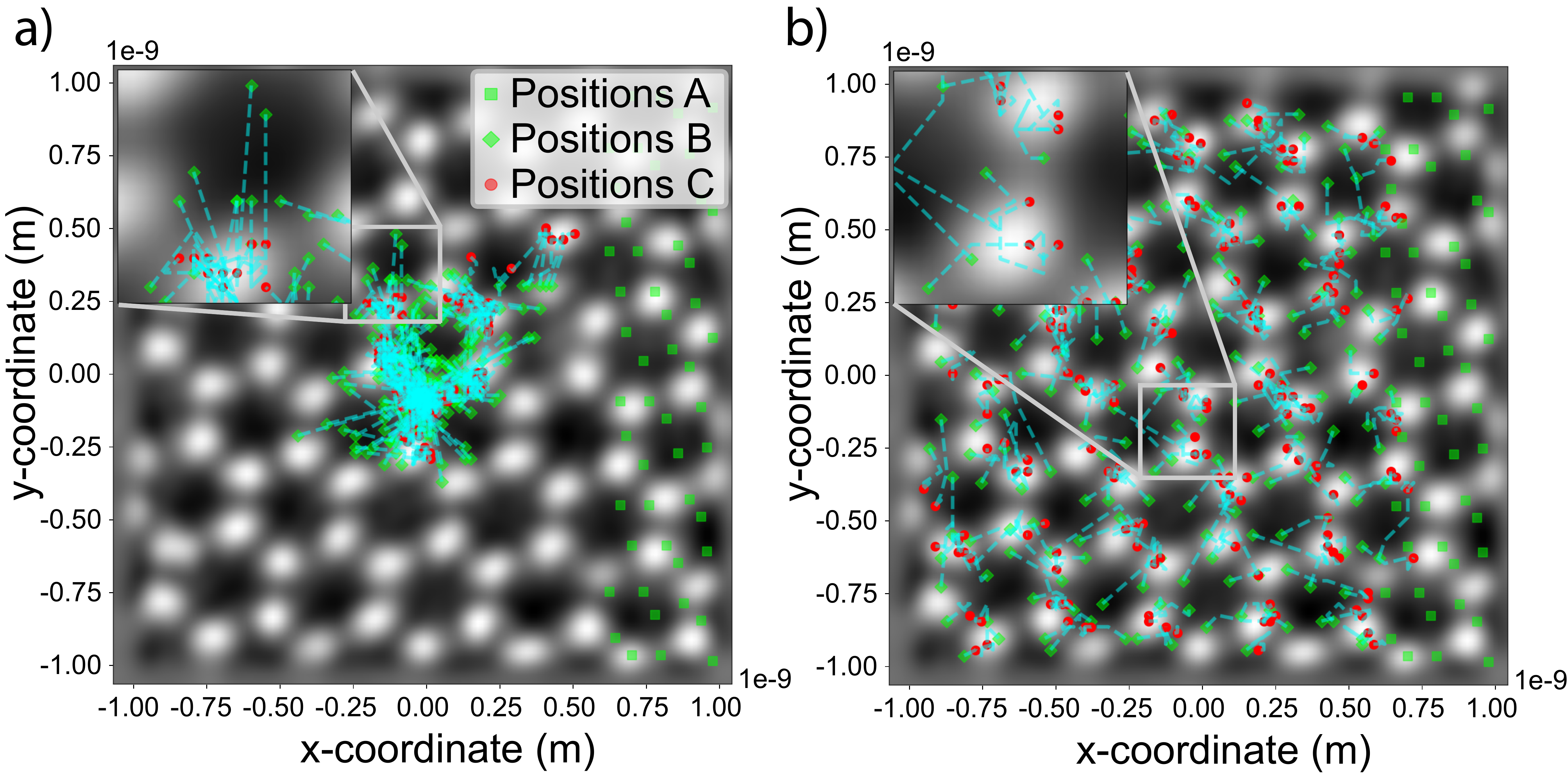}
\centering
\caption{Fine tuning of a policy with RL that a) has not been initialized and b) has been initialized via supervised learning. Positions A indicate the scan positions of the first sub-sequence $\vec{\textbf{R}}_{P_0}$ that is provided to the RNN as part of the initial input. Positions B and C are the scan positions of all predicted sub-sequences at iteration $0$ and 10,000, respectively. The trajectories they form during the optimization process are indicated by a dashed blue lines.}
\label{fig:comparisoninitialization}
\end{figure}

While training in RL can be performed with a policy whose parameters are arbitrarily initialized, this is not ideal. Having an adequate initial guess of the policy and using RL subsequently to only fine tune the policy is a much easier problem to solve. A quasi-random Halton sequence \cite{halton1964algorithm} with equally spaced probe positions is a reasonable initialization. Pre-training of the parameterized policy for the RL model can then be performed by supervised learning applied on the RNN such that the discrepancy between the predicted scan positions $\vec{\textbf{R}}_{P_t} = \vec{\boldsymbol{\mu}}_{\boldsymbol{\theta}}(\boldsymbol{h}_t)$ and the scan positions of the initialization sequence $\vec{\textbf{R}}^{\text{init}}_{P_t}$ is minimized:
\begin{equation}
\label{eq:metric}
\mathcal{K}(\boldsymbol{\theta}) = \sum^T_t \Vert \vec{\textbf{R}}_{P_t} - \vec{\textbf{R}}^{\text{init}}_{P_t} \Vert^2_2.
\end{equation}
Figure \ref{fig:comparisoninitialization} illustrates the scan positions during the fine tuning of the policy through RL for the first 10,000 iterations when either a) a policy that has not been initialized via supervised learning or b) an initialized policy is used. While the scan positions in both cases converge to the atomic structure, the positions predicted by the non-initialized policy are distributed only within a small region of the field of view during the entire training.

\section{Settings}

The various settings used for ROP, the convolutional autoencoder, and the RNN can be found in Table~\ref{tab:settings}.

\label{app:table}
\begin{table}[H]
\centering
\caption{Settings for the potential reconstruction with ROP,  structure content compression with the convolutional autoencoder and prediction of the scan sequences with the RNN.}
\begin{tabular}{llc}
\hline 
\hline
ROP &   \\
\hline
acceleration voltage (kV)                    & $60$			 \\
convergence angle (mrad)                    & $33$			 \\
object dimension (px)                                    & $200$ 	   \\
real space pixel size (nm)                               & $0.0154$ 	 \\
diffraction pattern dimension (px)                                    & $64$ 	   \\
reciprocal space pixel size $\left(\text{nm}^{-1}\right)$ & $0.4759$   \\
scanning step size (nm)                        & $0.02$   \\
iterations                        & $5$   \\
step size $\alpha_{\text{ROP}}$                     & $5.25\ensuremath{\text{\textsc{e}}}2$  \\
batch size                     & $24$   \\
\hline
Conv. autoencoder &   \\
\hline
input dimension (px)                                    & $512$ 	   \\
pixel size (nm)                               & $0.0064$ 	 \\
encoder/decoder kernel sizes (px)                               & $[3,3,3,3,3,3]$ 	 \\
encoder/decoder kernel strides (px)                               & $[1,1,1,1,1,1]$ 	 \\
encoder output channels                                & $[16,32,64,128,256,512]$ 	 \\
decoder output channels                                & $[256,128,64,32,16,2]$ 	 \\
iterations                        & $100000$   \\
step size $\alpha_{\text{CAE}}$                     & $1\ensuremath{\text{\textsc{e}-}}5$   \\
batch size                     & $24$   \\
\hline
RNN &   \\
\hline
sequence length                    & $250$			 \\
sub-sequence length                                    & $50$ 	   \\
hidden state $H_t$ size & $2048$   \\
stacked GRU layers                               & $2$ 	 \\
iterations                        & $20000$   \\
step size $\alpha_{\text{RNN}}$                     & $1\ensuremath{\text{\textsc{e}-}}6$   \\
batch size                     & $24$   \\
\hline 
\hline
\end{tabular}
\label{tab:settings}
\end{table}

\section{Examples}

\begin{figure*}
\includegraphics[width=1.0\linewidth]{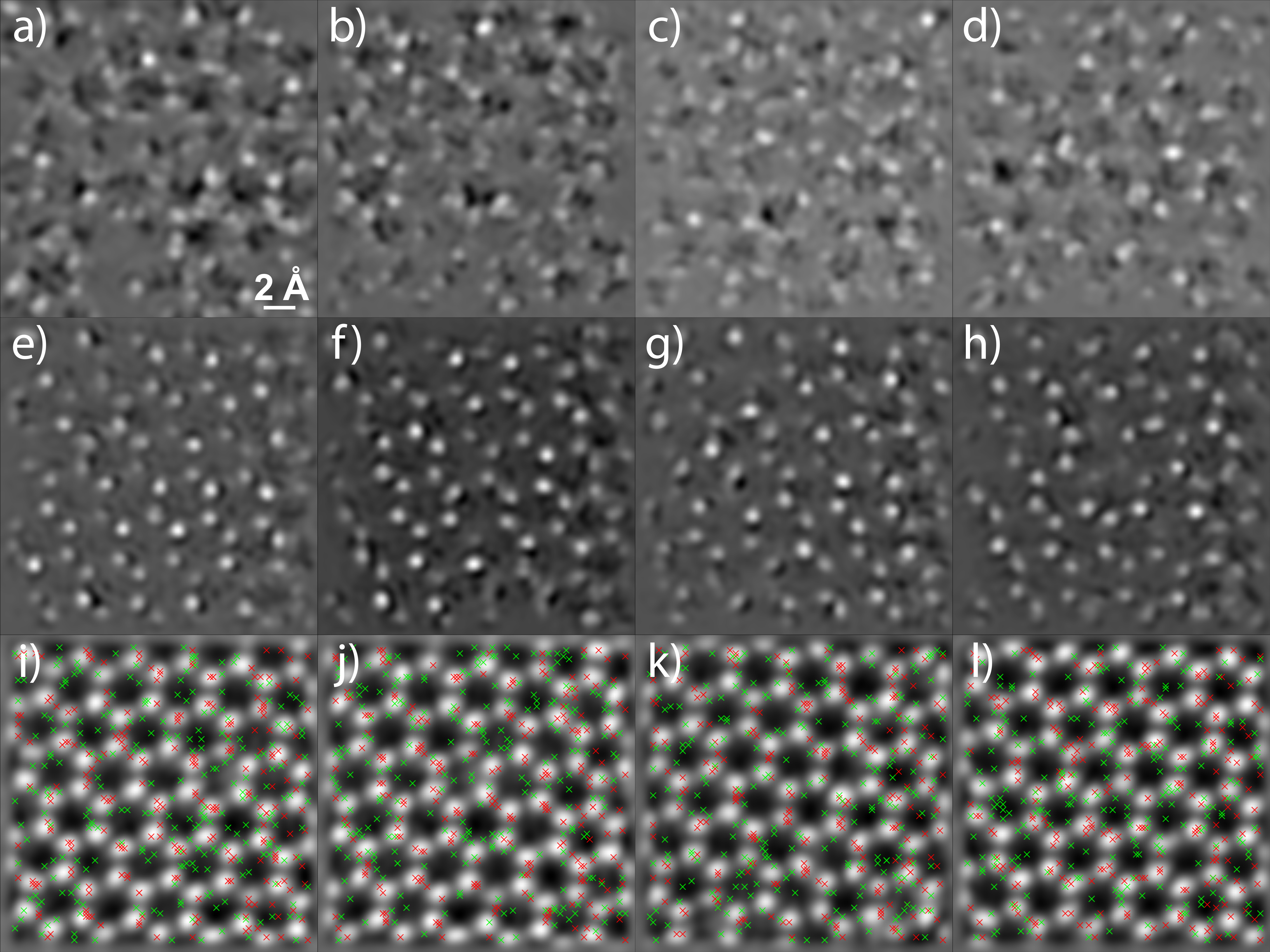}
\centering
\caption{Ptychographic reconstructions of different MoS$_2$ data sets and with different scanning procedures. Reconstruction from $250$ diffraction patterns of a data set that correspond to scan positions which follow a)-d) a random sequence and e)-h) an adaptively predicted sequence. e)-h) Ground truth reconstruction of the full data set with 10,000 diffraction patterns shown with the scan positions used for the corresponding reconstructions a)-d) in green and e)-h) in red.}
\label{fig:result_examples}
\end{figure*}

Several different exemplary reconstructions using adaptive scanning can be seen in Figure~\ref{fig:result_examples}.

\label{app:examples}

\clearpage

%===============================================================

\bibliography{main}% Produces the bibliography via BibTeX.

\end{document}